\documentstyle[preprint,eqsecnum,aps]{revtex}
\begin{document}
\draft
\title{Confinement and scaling in deep inelastic scattering}
\author{S. A. Gurvitz}
\address{Department of Particle Physics, Weizmann Institute of
         Science, Rehovot 76100, Israel\\
and TRIUMF, Vancouver, B.C., Canada V6T\ 2A3}
\maketitle
\begin{abstract}
We show that parton confinement in the final state generates
large $1/Q^2$ corrections to Bjorken scaling, thus leaving
less room for the logarithmic corrections. In particular, the
$x$-scaling violations at large $x$ are entirely
described in terms of power corrections. For treatment of these
non-perturbative effects, we derive
a new expansion in powers of $1/Q^2$ for the structure function
that is free of infra-red singularities and which reduces corrections
to the leading term. The leading term represents
scattering from an off-mass-shell parton, which keeps the same
virtual mass in the final state. It is found that this quasi-free
term is a function of a new variable $\bar x$, which coincides with
the Bjorken variable $x$ for $Q^2\to\infty$.
The two variables are very different, however, at finite $Q^2$.
In particular, the variable $\bar x$ depends on the invariant mass of the
spectator particles. Analysis of the data at large $x$ shows
excellent scaling in the variable $\bar x$, and determines the
value of the diquark mass to be close to zero. $\bar x$-scaling
allows us to extract the structure function near the elastic threshold.
It is found to behave as $F_2\sim (1-x)^{3.7}$. Predictions for
the structure functions based on $\bar x$-scaling are made.
\end{abstract}

\section{Introduction}
Consider the inclusive scattering of a high energy electron,
$e+N\rightarrow e'+X$, from a nucleon of
mass $M$. The spacelike 4-momentum
transferred to the target is $q=(\nu,{\mbox{\boldmath $q$}})$. For an
unpolarized target the double-differential cross-section is determined by
two structure functions $W_1(Q^2, \nu )$ and $W_2(Q^2, \nu )$
(or $F_1=MW_1$, $F_2=\nu W_2$), where $Q^2={\mbox{\boldmath $q$}}^2 -\nu^2$
These structure functions are given by
the imaginary part of the forward Compton amplitude of the virtual photon with
4-momentum $q$ (Fig. \ref{fig1}), where the nucleon vertex $\Gamma$ is
shown in Fig. \ref{fig2}.
The first diagram in Fig. \ref{fig1} corresponds to the Impulse
Approximation (IA), and the second diagram describes the
Final State Interaction (FSI) of a
struck quark with spectator quarks and gluons (these
are shown explicitly in Fig. \ref{fig2}). The IA term
is expected to become dominant in the structure function
for $Q^2\gtrsim 10$ (GeV/c)$^2$.
As a result $F_1$ and $F_2$ turn out to be functions
of the Bjorken variable
$x=Q^2/2M\nu$, i.e. $F_i(x,Q^2)\to F_i(x)$, where $F_i(x)$
is directly related to the parton distribution $\tilde{\rm q}(x)$
(the Bjorken scaling), as for instance
\begin{equation}
\nu W_2(Q^2,\nu )=F_2(x,Q^2)\to\sum_i e_i^2x \tilde{\rm q}_i(x).
\label{a1}
\end{equation}

The existing data in fact, show considerable $Q^2$-dependence of
the structure functions, as for the proton structure
function from the BCDMS\cite{bcd} and SLAC\cite{slac} experiments shown
in Fig. \ref{fig3}. (The solid lines correspond to a
15 parameter fit\cite{nmc}).
Usually the scaling deviation of the structure functions is
attributed mainly to the logarithmic corrections from
gluon radiation, which is contained in the
first diagram in Fig. \ref{fig1}\cite{rob}.
Corrections $\sim 1/Q^2$ arising from the second diagram
in Fig. \ref{fig1} (higher twist terms) are usually considered as playing
a minor role in scaling violation, even at moderate $Q^2$.

This common disregard
of the FSI terms for $Q^2\gtrsim 10$ (GeV/c)$^2$ is very surprising,
especially in view of
parton confinement. At first sight, the confining
interaction of partons in the final state should influence the
structure function strongly. Consider for instance the example of
two {\em nonrelativistic} ``quarks" of mass $m$ interacting via a harmonic
oscillator potential\cite{greenb}. These quarks are never free
and therefore the system in the final state possesses a discrete spectrum.
As a result the structure function, $F(q,\nu )$, as a function of $\nu$, is
given by a sum of $\delta$ - functions.
Obviously, it looks very different from the structure
function obtained in the IA, which considers the struck parton as
a free particle in the final state. This paradox can be resolved by
introducing a scaling
variable $y$\cite{w}
\begin{equation}
y=-\frac{|{\mbox{\boldmath $q$}}|}{2}+\frac{m\nu}
{|{\mbox{\boldmath $q$}}|},
\label{a2}
\end{equation}
where $(m-y)/M$ is a non-relativistic analogue of the Bjorken variable $x$.
Then expanding the structure function  ${\cal F}(q,y )\equiv F(q,\nu )$
in powers of $1/q$, one finds in the limit $q\to\infty$ and $y=$const that
it becomes a smooth curve,
${\cal F}(q,y )\to {\cal F}_0(y )$, which coincides with
a free parton response\cite{greenb,gr}.
Although this result appears to confirm the
parton model picture, it does not imply that the interaction in the final
state is not important. The latter has been merely incorporated in
${\cal F}_0(y )$ by an appropriate choice of the scaling
variable $y$, which diminishes the contribution from higher-order ($\sim 1/q$)
correction terms. For instance, a different choice of the scaling variable
could result in very large or even singular corrections
to the structure function.

It is practically impossible to calculate the structure
function including FSI, except for a few simple nonrelativistic
models\cite{greenb,gr}. Therefore it is very important to find
an optimal expansion of the structure function of confined systems,
which reduces higher order corrections ($\sim 1/q$)
to the zero order (quasi-free) term.
For a non-relativistic case such an expansion, which leads to the scaling
variable $y$, Eq. (\ref{a2}), was proposed by Gersch, Rodriguez and
Smith\cite{grs}.
(This expansion was designed for weakly bound systems, but
appears in fact to be applicable to confined systems as well\cite{gr}).
Unfortunately, the situation in the relativistic case looks very different,
and no simple extrapolation of the non-relativistic results seems to be
possible.

In this paper we attempt an optimal expansion for the
{\em relativistic$\;$}structure function, that
can be applied to confined systems, i.e., it is free of
infra-red singularites,
and diminishes the contribution from FSI. Then the zero order (quasi-free)
term effectively incorporates effects of {\em confining} FSI,
and thus can be considered
a good approximation for the structure function, valid also for
{\em nonasymptotic} $Q^2$. Such a quasi-free approximation leads to
scaling of the structure function in a new scaling variable.
Finally, we perform the analysis of data in terms of this scaling variable
and compare the results with a standard approach.

The plan is as follows: A relativistic expansion
of the structure function in powers of $1/Q^2$ is derived in Section 2.
It is shown that infra-red singularities generated by confining FSI
are eliminated. The leading, quasi-free term is
discussed in Section 3. We demonstrate there that the structure
function is a function of a new scaling variable. The evaluation of
the first correction term for the linear-rising potential is presented in
Section 4. An analysis of data and predictions for new experiments
are given in Section 5. The last Section is summary.

\section{Relativistic structure function}

Consider the nucleon structure function $W$ given by the imaginary part
of the forward Compton amplitude, Fig. \ref{fig1}.
For the sake of simplicity we
take all the consituents and the virtual photon as scalar particles.
Here $P=(M,0)$ is the 4-momentum of a target (in the laboratory frame),
$P-p$ is the 4-momentum of the struck parton and
$p=(p_0, {\mbox{\boldmath $p$}})$ is the total 4-momentum of all other
constituents (quarks and gluons) to which we refer as the ``spectator".
The vertex $\Gamma$ is a sum of all possible diagrams describing the virtual
nucleon disintegration into quarks and gluons, Fig. \ref{fig2}.

The scattering amplitude (the square block in Fig. \ref{fig1})
satisfies a Bethe-Salpeter equation, shown schematically in
Fig. \ref{fig4},
\begin{equation}
T=V+VG_0T,
\label{b1}
\end{equation}
where the Green's function $G_0$ can be written as
\begin{equation}
\langle p|G_0({\cal P})|p'\rangle = ig_s(p)g_0({\cal P}-p)\delta (p-p').
\label{b2}
\end{equation}
Here $g_s(p)$ is the spectator Green's function, and $g_0({\cal P}-p)$ is a
Green's function for a struck quark :
\begin{equation}
g_0({\cal P}-p)=\frac{1}{({\cal P}-p)^2-m^2+i\epsilon },
\label{b2a}
\end{equation}
where ${\cal P}$ is the total 4-momentum of the system
(${\cal P}=P+q$ for the kinematics in Fig. \ref{fig1}) and $m$ is the struck
parton mass. The driving (interaction) term $V$ in Eq. (\ref{b1})
is a sum of all irreducible diagrams which do not include
the struck quark and the spectator in an intermediate
state. Since quarks are confined at large distances, this term is singular
for $(p-p')^2\rightarrow 0$ (for instance,
$\langle p |V|p'\rangle\sim (p-p')^{-4}$ in the case of linear-rising
confinement).

The corresponding Bethe-Salpeter equation for the
vertex function $\Gamma$ is obtained from Eq. (\ref{b1}) by taking the
limit ${\cal P}^2\rightarrow P^2= M^2$.
Since the  amplitude $T$  factorizes near the nucleon pole
\begin{equation}
T({\cal P},p,p')\rightarrow \frac{\Gamma (P-p,p)\Gamma (P-p',p')}
{{\cal P}^2-M^2}
\label{b3}
\end{equation}
one finds the following equation
\begin{equation}
\Gamma =V G_0(P)\Gamma,
\label{b4}
\end{equation}
which is shown schematically in Fig. \ref{fig5}.

By introducing the interacting (full) Green's function
\begin{equation}
G=G_0+G_0VG_0+\cdots = \frac{1}{G_0^{-1}-V}
\label{b5}
\end{equation}
we can represent the structure function $W(Q^2,\nu )$ as
\begin{equation}
W(Q^2,\nu )=\frac{1}{\pi}
{\rm {Im}}\int \Phi (P,p)\langle p\big \vert G(P+q)\big \vert p'\rangle
\Phi (P,p')\frac{d^4pd^4p'}{(2\pi )^8},
\label{b6}
\end{equation}
where
\begin{equation}
\Phi (P,p)=\frac{\Gamma (P-p,p)}{(P-p)^2-m^2}\equiv g_0(P-p) \Gamma (P-p,p)
\label{b7}
\end{equation}
is the relativistic bound state wave function.

The IA of the structure function $W$
corresponds to $G\to G_0$ in Eq. (\ref{b6}). This approximation can be
applied if the contribution from the interaction $V$ is small.
However, this may not be
the case, in particular because of the infra-red singularity in $V$.
The singularity can generate large corrections from the higher
order terms of the expansion (\ref{b5}). Yet, it does
not imply that the correction terms would remain large in any other
expansions of the Green's function $G$. Indeed, the bound state
wave function is generated by the same driving term $V$, Eq. (\ref{b4}), so
that the singular part of $V$ can be compensated by some part of the mass and
kinetic energy terms of in the full Green's function $G=[G_0^{-1}-V]^{-1}$,
when the latter is substituted into Eq. (\ref{b6}).
In fact, one finds from Eqs. (\ref{b4}), (\ref{b7})
the following relation
\begin{equation}
[V-G_0^{-1}(P)]g_s|\Phi \rangle =0,
\label{bb7}
\end{equation}
which in nonrelativistic limit corresponds to
cancellations between the binding potential and the
kinetic and binding energy terms in the
Schroedinger equation.

Eq. (\ref{bb7}) suggests to expand the total Green's function in (\ref{b6})
in powers of the operator $h\equiv h(P)=V-G_0^{-1}(P)$
instead of the IA expansion (\ref{b5}), in powers of $V$.
We thus obtain
\begin{equation}
G=\tilde G+\tilde Gh\tilde G+\tilde Gh\tilde Gh\tilde G+\cdots,
\label{b9}
\end{equation}
where $\tilde G^{-1}\equiv\tilde G^{-1}({\cal P},P)= G^{-1}({\cal P})-
G^{-1}(P)$. It can be rewritten explicitly as
\begin{equation}
\langle p|\tilde G^{-1}({\cal P},P)|p' \rangle  =
\langle p|V(P)-V({\cal P})|p' \rangle+g_s^{-1}(p)[({\cal P}-p)^2-
(P-p)^2]\delta (p-p')
\label{bb9}
\end{equation}
One notes that in a general, the driving term $V$ is not local,
and therefore it may depend on the total 4-momentum ${\cal P}$
of the whole system (Fig. \ref{fig4}). However,
for a local interaction,
$\langle p|V({\cal P})|p' \rangle\equiv V(p-p')$, so that the terms
containing $V$ in Eq. (\ref{bb9}) cancel. Then
\begin{equation}
\langle p|\tilde G({\cal P},P)|p'\rangle = i\ g_s(p)\tilde g({\cal P}-p,
P-p)\delta (p-p'),
\label{b10}
\end{equation}
where
\begin{equation}
\tilde g({\cal P}-p,P-p)=\frac{1}
{({\cal P}-p)^2-(P-p)^2+i\epsilon}
\label{b11}
\end{equation}
is a modified, quasi-free, Green's function of the struck parton with
4-momentum ${\cal P}-p$. Unlike the Green's function $g_0$,
Eq. (\ref{b2a}), the modified Green's function
$\tilde g$, as well as $\tilde G$, depends
on the target 4-momentum $P$, which is related
to the entire interacting system. The appearance of such an
additional parameter
is not surprising, since $\tilde g$ has been designed to approximate the
{\em interacting} Green's function.  Note that
the pole in $\tilde g$ does not appear at the quark mass $({\cal P}-p)^2=m^2$
as in the free Green's function $g_0$, but at an off-shell point,
$({\cal P}-p)^2=(P-p)^2$.

Substituting Eqs. (\ref{b9}), (\ref{b10})
into Eq. (\ref{b6}) we obtain the
structure function $W$ in an expansion
in powers of $\tilde G=g_s\tilde g$ for ${\cal P}=P+q$. Consider the limit
$Q^2\rightarrow\infty$ and $x=Q^2/2m\nu$ = const. Then
$\nu=Q^2/2Mx\sim Q^2$,  $\; \nu/|{\mbox{\boldmath $q$}}|\to 1$, and
\begin{eqnarray}
\tilde g^{-1}(P+q-p,P-p) & = &
(P-p+q)^2-(P-p)^2+i\epsilon\nonumber\\
& = & 2(M-p_0)\nu +2{\mbox{\boldmath $p$}}\cdot
{\mbox{\boldmath $q$}}-Q^2+i\epsilon
\sim Q^2
\label{b12}
\end{eqnarray}
Therefore Eq. (\ref{b9})
represents in fact an expansion in powers of $1/Q^2$, and
\begin{equation}
F = \nu W =\frac{\nu}{\pi} {\rm {Im}}
\langle\Phi |\tilde G+\tilde Gh\tilde G+\cdots |\Phi\rangle
={\cal F}_0+\frac{{\cal F}_1}{Q^2}+\frac{{\cal F}_2}{Q^4}+\cdots
\label{b13}
\end{equation}
Each term of this expansion can be represented by a modified
Feynman diagram. Take for instance the first two terms
\begin{mathletters}
\label{ab13}
\begin{equation}
{\cal F}_0 = \frac{\nu}{\pi}{\rm {Im}} \int i\frac{d^4p}{(2\pi)^4}
\frac{\Phi^2(P,p)g^2_s(p)}{(P-p+q)^2-(P-p)^2+i\epsilon}
\label{ab13a}
\end{equation}
\begin{equation}
\frac{{\cal F}_1}{Q^2}  =  \frac{\nu}{\pi}{\rm {Im}}
\int i\frac{d^4pd^4p'}{(2\pi)^8}
\frac{\Phi (P,p)g_s(p)h(P,p,p')g_s(p')\Phi (P,p')}
{[(P-p+q)^2-(P-p)^2+i\epsilon ][(P-p'+q)^2-(P-p')^2+i\epsilon ]}
\label{ab13b}
\end{equation}
\end{mathletters}
One can easily see that these terms correspond to the two
diagrams in Fig. \ref{fig6}, where the Feynman propagator $(k^2-m^2)^{-1}$
for the struck parton with 4-momentum $k$ is replaced by
$[k^2-(k-q)^2]^{-1}$. We mark it with ``$\sim$".

Let us consider the first order term, ${\cal F}_1$, Eq. (\ref{ab13b}),
which involves the interaction $V$.
One gets from Eqs.(\ref{bb7}), (\ref{b10})
\begin{mathletters}
\label{bb13}
\begin{eqnarray}
h\tilde gg_s|\Phi\rangle & = & [h,\tilde g ]g_s|\Phi\rangle =
[V,\tilde g]g_s|\Phi\rangle\label{bb13a}\\
\langle\Phi|g_s\tilde g h & = & \langle\Phi|g_s[\tilde g, h]=
-\langle\Phi|g_s[V,\tilde g]
\label{bb13b}
\end{eqnarray}
\end{mathletters}
Therefore Eq. (\ref{ab13b}) can be rewritten as
\begin{equation}
{\cal F}_1=\frac{\nu}{2\pi} Q^2
\ {\rm {Im}}\ i\langle\Phi |g_s\left [[\tilde g,V],
\tilde g\right ]g_s|\Phi\rangle
\label{b14}
\end{equation}
where the interaction enters through the double commutator of
$V$ and the Green's function $\tilde g$.
Using Eqs. (\ref{b11}), (\ref{b12}) we can write Eq. (\ref{b14})
explicitly as
\begin{equation}
{\cal F}_1=\nu Q^2
\ {\rm {Im}}\ \int i\frac{d^4pd^4p'}{(2\pi )^9}
\frac{\Phi (P,p)g_s(p)
\left (2q(p-p')\right )^2V(p-p')g_s(p')\Phi (P,p')}
{[2(M-p_0)\nu +2{\mbox{\boldmath $p$}}\cdot
{\mbox{\boldmath $q$}}-Q^2+i\epsilon]^2
[2(M-p'_0)\nu +2{\mbox{\boldmath $p'$}}\cdot
{\mbox{\boldmath $q$}}-Q^2+i\epsilon]^2}
\label{b15}
\end{equation}
Here we find the factor $\left (q(p-p')\right )^2$ in front of $V$,
which removes the infra-red singularity, $V\sim 1/(p-p')^4$,
in ${\cal F}_1$. In fact, this factor
reduces the contribution of ${\cal F}_1$ to the structure
function even for a non-singular interaction, provided that the scattering
amplitude of a high-momentum quark peaks in the forward
direction. One can show that the same procedure removes the
infra-red singularity also in higher orders in the expansion
(\ref{b13}) for $F$. A direct evaluation of ${\cal F}_1$
for the case of a heavy spectator is given in Sec. 4, where we
explicitly demonstrate that this term is small
compared to ${\cal F}_0$, even at moderate $Q^2$. Now we
concentrate on the first term, ${\cal F}_0$, in the expansion
(\ref{b13}).

\section{The leading term}

\subsection{New scaling variable.}

Since our expansion (\ref{b13}) minimizes the first order correction
term, the structure function $F$ can be well approximated
by the first term,
${\cal F}_0$, Eq. (\ref{ab13a}), which corresponds to
the first graph in Fig. \ref{fig6}. Actually, this graph
represents an infinite
sum of diagrams corresponding to gluon and quark emission, Fig. \ref{fig7},
which results in logarithmic corrections to Bjorken scaling and
has been studied in great detail\cite{rob}.
Our treatment of these processes is not different from the standard
approaches except for the modified
propagator $\tilde g$, marked  by ``$\sim$" in Figs. \ref{fig6}, \ref{fig7}.
Let us analyze the consequences of this modification. As an example we
consider the two diagrams shown in
Fig. \ref{fig7}. The first graph corresponds
to a process with no gluon emission.
For simplicity we take the spectator quarks to be a
diquark of mass $m_s$, thus
\begin{equation}
g_s(p)=\frac{1}{p^2-m_s^2+i\epsilon}
\label{bb16}
\end{equation}
Substituting Eq. (\ref{bb16}) into Eq. (\ref{ab13a})
one finds
\begin{equation}
{\cal F}_0^{(0)}=
\nu \int \frac{d^4p}{(2\pi )^3}|\Phi^{(0)} (P,p)|^2\delta (p^2-m_s^2)\delta
[(P-p+q)^2-(P-p)^2],
\label{b16}
\end{equation}
where $\Phi^{(0)}$ is a component of the proton wave function, Eq. (\ref{b7}),
with the vertex $\Gamma$ replaced by $\Gamma^{(0)}$,
corresponding to nucleon disintegration with no gluon
emission (Fig. \ref{fig2}).
Integrating over $p_0$ and neglecting the contribution from negative
energy states (pair production) we get from the first $\delta$-function:
$p_0=E_p=(m_s^2+{\mbox{\boldmath $p$}}^2)^{1/2}$. Then using
(\ref{b12}) we can rewrite Eq. (\ref{b16}) as
\begin{eqnarray}
{\cal F}_0^{(0)} & = & \nu \int\frac{d^3p}{(2\pi )^3}
|\tilde\phi^{(0)} (|{\mbox{\boldmath $p$}}|)|^2
\delta \left [(M-E_p)\nu +
{\mbox{\boldmath $p$}}\cdot{\mbox{\boldmath $q$}}-Q^2/2\right ]
=\frac{\nu }{|{\mbox{\boldmath $q$}}|}\int_{|\tilde y|}^{\infty}
\frac{pdp}{(2\pi )^2}|\tilde\phi^{(0)}(|{\mbox{\boldmath $p$}}|)|^2
\nonumber\\
\noalign{\vskip7pt}
& = & \frac{\nu }{|{\mbox{\boldmath $q$}}|}\int
\frac{d^3p}{(2\pi )^3}|\tilde\phi^{(0)}(|{\mbox{\boldmath $p$}}|)|^2
\delta(p_z+\tilde y),
\label{b17}
\end{eqnarray}
where $\tilde\phi^{(0)} (|{\mbox{\boldmath $p$}}|)=\Phi(P,E_p,
{\mbox{\boldmath $p$}})/2E_p^{1/2}$.
The variable $\tilde y$ is the minimal momentum of the struck quark,
$-{\mbox{\boldmath $p$}}_{min}=
\tilde y{\mbox{\boldmath $q$}}/|{\mbox{\boldmath $q$}}|$,
obtained from the equation
\begin{equation}
2\left (M-\sqrt{\tilde y^2+m_s^2}\right )\nu -2\tilde y
|{\mbox{\boldmath $q$}}|-Q^2=0.
\label{b18}
\end{equation}
The latter corresponds to energy conservation, given by the
$\delta$-function in Eq. (\ref{b17}), when the struck quark is
equally off-mass shell before and after
the virtual photon absorption. Solving Eq. (\ref{b18}) and
using $\nu =Q^2/2Mx$, $\;{\mbox{\boldmath $q$}}^2=Q^2+\nu^2$
we obtain
\begin{equation}
\tilde y(x,Q^2)/M=\frac{(1-x)^2-(m_s/M)^2}{\sqrt{(1-x)^2+4m_s^2x^2/Q^2}+
\sqrt{(1-x)^2+4M^2x^2(1-x)^2/Q^2}}
\label{b19}
\end{equation}
It follows from Eq. (\ref{b17}) that
$(|{\mbox{\boldmath $q$}}|/\nu ){\cal F}_0^{(0)}
\equiv f(\tilde y)$ is a function of the scaling variable
$\tilde y$ only. In the nonrelativistic limit and for zero binding energy,
this variable coincides with the West scaling variable $y$,
Eq. (\ref{a2}).

Eq. (\ref{b16}) for ${\cal F}_0^{(0)}$ can be rewritten
in terms of light-cone variables:
$p\equiv (p_+,p_-,{\mbox{\boldmath $p$}}_{\perp})$,
where $p_{\pm}=p_0\pm p_z$. Here the negative
$z$-axis has been chosen along the virtual photon direction, so that
$q_{\pm}=\nu\mp|{\mbox{\boldmath $q$}}|$, and ${\mbox{\boldmath $q$}}_{\perp}=
0$. Introducing the light-cone fractions $z=(P_+-p_+)/P_+$ and $\xi= -q_+/P_+$,
where $P_\pm=M$ is the target light-cone momentum, and integrating over
$p_-$, one obtains
\begin{equation}
{\cal F}_0^{(0)}=
\nu \int \frac{d^2p_{\perp}dz}{2(2\pi )^3}
\frac{|\bar\Phi^{(0)} (p_{\perp},z)|^2}{(1-z)(z-\xi )}
\delta\left (M^2+Mq_--\frac{{\bar m}^2+p_{\perp}^2}{z-\xi}-
\frac{m_s^2+p_{\perp}^2}{1-z}\right )
\label{bc1}
\end{equation}
where $Mq_-=Q^2/\xi$ and
\begin{equation}
{\bar m}^2=(P-p)^2=zM^2-\frac{zm_s^2+p_{\perp}^2}{1-z}
\label{bc2}
\end{equation}

The expression (\ref{bc1}) for ${\cal F}_0^{(0)}$ clearly
corresponds to the first diagram
in Fig. \ref{fig8}, calculated according the rules of light-cone
perturbation theory, where the struck quark mass equals $\bar m$.
Similar to the previous calculations one finds after
integration over ${\mbox{\boldmath $p$}}_{\perp}$ and $z$ in (\ref{bc1}) that
the structure function depends only on a single (scaling) variable $\bar x$.
The latter is the value of the light-cone fraction $z$, corresponding to
the zero of the $\delta$-function argument in Eq. (\ref{bc1})
for $p_{\perp}=0$. One finds
\begin{equation}
\bar x=\frac{x+\sqrt{1+4M^2x^2/Q^2}-
\sqrt{(1-x)^2+4m_s^2x^2/Q^2}}
{1+\sqrt{1+ 4M^2x^2/Q^2}}
\label{b21}
\end{equation}
Since $\bar x$ corresponds to the light-cone fraction of the
off-shell struck quark with 4-momentum $P-p$ (Fig. \ref{fig7}), where
$p=(\sqrt{m_s^2+\tilde y^2},
-\tilde y{\mbox{\boldmath $q$}}/|{\mbox{\boldmath $q$}}|)$,
one can easily verify that $\bar x$ is related to $\tilde y$ by
\begin{equation}
\bar x=1-\frac{\sqrt{m_s^2+\tilde y^2}+\tilde y}{M}
\label{bc3}
\end{equation}

It is interesting to compare $\bar x$ with the Nachtmann\cite{nacht}
scaling variable, $\xi$
\begin{equation}
\xi =\frac{2x}{1+\sqrt{1+4M^2x^2/Q^2}},
\label{b22}
\end{equation}
which corresponds to the light-cone fraction of the
on-mass-shell struck parton of zero mass\cite{iof}. The denominator
of this expression takes into account the kinematical target mass
corresctions to the $x$-scaling. Similar target mass effects
are accounted for in Eq. (\ref{b21}) for the scaling variable $\bar x$.
It is not surprising, since no assumptions have been made for
the value of the target mass $M$ in Eqs. (\ref{b17}), (\ref{bc1}). However,
$\bar x$ includes also dynamical corrections to the $x$-scaling,
which are not present in the variable $\xi$. These are taken
into account by the last term
in the numerator of Eq. (\ref{b21}).  It can be seen in a more pronounced
way if we consider large $Q^2$ limit. Then using  Eqs. (\ref{b21}), (\ref{b22})
one easily finds that $\bar x$ and $\xi$-variables are related by
\begin{equation}
\bar x \simeq \xi +\frac{M^2x^2}{Q^2} -\frac{m_s^2x^2}{(1-x)Q^2}
\label{b22p}
\end{equation}
Thus $\bar x\to\xi\to x$ for $Q^2\to\infty$. However, the term
$m_s^2x^2/(1-x)Q^2$ makes $\bar x$ and $\xi$
be quite different for finite $Q^2$, in particular for high $x$.
As an example we plot the variables $\xi$ and $\bar x$ as
functions of $Q^2$ for $x=0.75$ in Fig. \ref{fig9},
for a spectator mass $m_s=M$.
One finds that $\bar x$ approaches the Bjorken variable
$x$ (the dotted line) much more slowly than the Nachtmann variable $\xi$, and
the difference between $\bar x$ and $\xi$ is
appreciable even for rather large $Q^2$.
For small $x$, however, the variable $\bar x$ is
close to $\xi$  or $x$, unless the spectator mass is not very large.

\subsection{Gluon radiation.}

Consider the second diagram in Fig. \ref{fig7},
which describes gluon emission.
Its contribution to the structure function can be written as
\begin{equation}
{\cal F}_0^{(1)}=
\nu \int \frac{d^4pd^4p'}{(2\pi )^8}|\Phi^{(1)} (P,p',p-p')|^2
\delta (p'^2-m_s^2)\delta (p-p')^2 \delta [(P-p+q)^2-(P-p)^2]
\label{b23}
\end{equation}
where $\Phi^{(1)}$ is the component of the proton wave function corresponding
to the vertex $\Gamma^{(1)}$ in Fig. \ref{fig2}.
It is related to $\Phi^{(0)}$ by
\begin{equation}
|\Phi^{(1)} (P,p',p-p')|^2=\frac{4\pi\alpha_s|\Phi^{(0)} (P,p')|^2}
{(p-p')^2-m^2},
\label{b23a}
\end{equation}
where $\alpha_s$ is the QCD running coupling constant. After integration over
the $p_0$ and ${p'}_0$ we can rewrite ${\cal F}_0^{(1)}$ in the
form of Eq. (\ref{b17}):
\begin{equation}
\frac{|{\mbox{\boldmath $q$}}|}{\nu}
{\cal F}_0^{(1)}=\int\frac{d^3pd^3p'}{(2\pi )^6}
|\tilde\phi^{(1)} ({\mbox{\boldmath $p$}},{\mbox{\boldmath $p'$}})|^2
\delta [p_z+ \tilde y(m_s)].
\label{b24}
\end{equation}
Here $\tilde y(m_s)$ is given by Eq. (3.5), where $m_s$ is now
the invariant mass of the diquark-gluon system: $m_s^2= p^2$.
In terms of light-cone variables (see Fig. \ref{fig8}) it can
be written as
\begin{equation}
m_s^2=m_d^2+(1-z)\frac{
({\mbox{\boldmath $p'$}}_{\perp}-{\mbox{\boldmath $p$}}_{\perp})^2}
{z'-z}+(z'-z)\frac{m_d^2+{\mbox{\boldmath $p'$}}^2}{1-z'}+
{\mbox{\boldmath $p'$}}_{\perp}^2-{\mbox{\boldmath $p$}}_{\perp}^2,
\label{b25}
\end{equation}
where $m_d$ is the diquark mass. Notice  that $1\geq z'\geq z$.
Using Eq. (\ref{b25}) we rewrite Eq. (\ref{b23}) in terms of light-cone
variables
\begin{equation}
{\cal F}_0^{(1)}=
\nu \int \frac{d^2p_{\perp}dzd^2p'_{\perp}dz'}{4(2\pi )^6}
\frac{|\bar\Phi^{(1)} (p_{\perp},z,p'_{\perp},z')|^2}{(1-z')(z'-z)(z-\xi )}
\delta\left (M^2+Mq_--\frac{{\bar m}^2+p_{\perp}^2}{z-\xi}-
\frac{m_s^2+p_{\perp}^2}{1-z}\right ),
\label{b26}
\end{equation}
where $\bar m^2 =(P-p)^2$, Eq. (\ref{bc2}).

Eq. (\ref{b26}) describes the lowest order gluon emission contribution to the
structure function, ${\cal F}_0$.
The total contribution of gluon emission to ${\cal F}_0$ corresponds to the
sum of all ladder diagrams. It can be done using the same procedure as for
instance in\cite{pol,dok}. Our modification consists only of
replacement of the struck quark mass $m$ by $\bar m$,  Eq. (\ref{bc2}).
Eventually, it will modify the evolution
equation\cite{alt} by the replacement of the Bjorken
scaling variable $x$ by $\bar x$, Eq. (\ref{b21}),
for $m_s$ given by Eq. (\ref{b25}).

\subsection{Approximation.}

A considerable simplification can be achieved if we approximate
$m_s$ as an effective spectator mass
depends only on external momenta. Since $z\to\bar x\simeq x$ and
$z'\sim z$, one gets from Eq. (\ref{b25})
\begin{equation}
m_s^2\simeq m_d^2+C(x,Q^2)(1-x)
\label{b27}
\end{equation}
In this case the off-shell mass $\bar m$ of the struck quark, Eq. (\ref{bc2}),
becomes independent
of $z',{\mbox{\boldmath $p'$}}_{\perp}$. This allows us to integrate
over $z',{\mbox{\boldmath $p'$}}_{\perp}$ in Eq. (\ref{b26}), thus
reducing it to the form of Eq. (\ref{bc1})
(or (\ref{b17})). Finally one can sum all the ladder diagrams,
${\cal F}_0^{(i)}$ such that
$|\bar\Phi^{(0)}(p_{\perp},z)|^2$ in Eq. (\ref{bc1}) is replaced by
$u(p_{\perp},z,Q^2)$, which is a probability
to find a struck quark with
momentum $p$ inside the nucleon, and a spectator with any number of
gluons. The latter gives rise to logarithmic corrections to scaling.

Eq. (\ref{b27}) for the invariant spectator mass looks quite appealing
apart from its relation to Eq. (\ref{b25}). Indeed, $x=1$ corresponds to
elastic scattering, when no gluons are emitted. Therefore in this
case the spectator is represented by a diquark. When
$x$ decreases, gluons are emitted and $m^2_s$ increases $\propto (1-x)$.
The coefficient $C(x,Q^2)$ in Eq. (\ref{b27})
determines the rate of increase of the spectator mass with $Q^2$ and $x$.
It can be found self-consistently from the evolution
equation. However, when $x\sim 1$, one can take
$C(x,Q^2)\simeq C(1,Q^2)\simeq$ const, because of
$Q^2$-dependence of the spectator mass is less important than its
$x$-dependence near the elastic threshold.
Let us roughly estimate the value of $C$ by using the Weizs\"acker-Williams
or ``equivalent photon" approximation, utilized in Ref.\cite{jaf}
for derivation of the evolution equation. One finds from\cite{jaf} that
the light-cone fraction of the ``equivalent" gluon,
$z-z'$, (Fig. \ref{fig8}) is of order $\alpha_s \ln (Q^2/Q_0^2)$
in the region of large $x$.
However, ${\cal F}_0^{(1)}/{\cal F}_0^{(0)}$
is also about the same order of magnitude. Then, as follows from
Eq. (\ref{b25}), $C\sim\langle
({\mbox{\boldmath $p'$}}_{\perp}-{\mbox{\boldmath $p$}}_{\perp})^2
\rangle$, so that one could expect to find $C$ on the scale of (GeV)$^2$.

\section{Correction term}

Consider the first correction ${\cal F}_1/Q^2$, Eq. (\ref{b13}),
to the leading term ${\cal F}_0$.
In order to simplify the evaluation of
${\cal F}_1$, Eq. (\ref{b15}), we take the large spectator
mass limit: $m_s\gg |{\mbox{\boldmath $p$}}|$, so that
$g_s\to \delta (p_0-m_s)/2m_s$.
In the same limit one gets from Eq. (\ref{b18})
\begin{equation}
\tilde y\to y=-\frac{Q^2}{2|{\mbox{\boldmath $q$}}|}+
\frac{m^*\nu}{|{\mbox{\boldmath $q$}}|}
\label{c1}
\end{equation}
where $m^*=M-m_s$, and also
the Green's function $\tilde g$,  Eq. (\ref{b12}), reads
\begin{equation}
\tilde g=\frac{1}{ 2|{\mbox{\boldmath $q$}}|(p_z+y
+i\epsilon)}.
\label{c2}
\end{equation}
Then the first order correction term ${\cal F}_1$, Eq. (\ref{b15}), becomes
\begin{equation}
{\cal F}_1=-{\cal A}_1
\ {\rm {Im}}\ \frac{1}{2\pi}\int\frac{d^3pd^3p'}{(2\pi )^6}\tilde\phi
({\mbox{\boldmath $p$}})
\frac{(\hat{\mbox{\boldmath $q$}}\cdot ({\mbox{\boldmath $p$}}-
{\mbox{\boldmath $p'$}}))^2V({\mbox{\boldmath $p$}}-{\mbox{\boldmath $p'$}})}
{(p_z+y+i\epsilon )^2 (p'_z+y+i\epsilon )^2}
\tilde\phi ({\mbox{\boldmath $p'$}})
\label{c3}
\end{equation}
where $\hat{\mbox{\boldmath $q$}}={\mbox{\boldmath $q$}}/
|{\mbox{\boldmath $q$}}|$, and
\begin{equation}
{\cal A}_1 =\frac{\nu^2}{{\mbox{\boldmath $q$}^2}}
\frac{Q^2}{4m_s\nu}=\frac{Mx}
{2m_s(1+4M^2x^2/Q^2)}\to \frac{Mx}{2m_s},
\;\;\;\; {\mbox{for $Q^2\to \infty$}}
\label{c4}
\end{equation}

It is convenient to evaluate the terms ${\cal F}_i$ in (\ref{b13}) using
coordinate representation. Substituting
\begin{equation}
\tilde\phi ({\mbox{\boldmath $p$}})=\int e^{i{\mbox{\boldmath $p$}}\cdot
{\mbox{\boldmath $r$}}}\phi ({\mbox{\boldmath $r$}})d^3r,
\;\;\;\;\;\;\;\;V({\mbox{\boldmath $p$}}-{\mbox{\boldmath $p'$}})=
\int e^{i({\mbox{\boldmath $p$}}-{\mbox{\boldmath $p'$}})\cdot
{\mbox{\boldmath $r$}}}v({\mbox{\boldmath $r$}})
d^3r
\label{c5}
\end{equation}
into Eqs. (\ref{b17}),(\ref{c3}) one obtains after some algebra\cite{rr}
the following expressions for the two first terms of the expansion
(\ref{b13})
\begin{mathletters}
\label{c6}
\begin{eqnarray}
{\cal F}_0 & = & \frac{{\cal A}_0}{2\pi}
\ \int_{-\infty}^{\infty}ds\exp (-iys)\int d^3r\phi({\mbox{\boldmath $r$}}-
s\hat{\mbox{\boldmath $q$}})\phi({\mbox{\boldmath $r$}})
\label{c6a}\\
\noalign{\vskip7pt}
{\cal F}_1 & = & i\frac{{\cal A}_1}{2\pi}
\ \int_{-\infty}^{\infty}ds\exp (-iys)\int d^3r\phi({\mbox{\boldmath $r$}}-
s\hat{\mbox{\boldmath $q$}})\phi({\mbox{\boldmath $r$}})\int_0^s d\sigma
[ v({\mbox{\boldmath $r$}}-\sigma \hat{\mbox{\boldmath $q$}})-
v({\mbox{\boldmath $r$}})]
\label{c6b}
\end{eqnarray}
\end{mathletters}
where ${\cal A}_0=\nu/|{\mbox{\boldmath $q$}}|\to 1$ for $Q^2\to\infty$.
These expressions are of the same form
(up to the coefficients ${\cal A}_i$) as
the first two terms of the nonrelativistic $1/|{\mbox{\boldmath $q$}}|$
expansion of the structure function\cite{grs}. One checks that
the same corerspondence also holds for higher order terms
${\cal F}_i$ in the expansion (\ref{b13}). Explicit analytical evaluations
of the first three terms in the nonrelativistic  $1/|{\mbox{\boldmath $q$}}|$
expansion\cite{grs} for the harmonic oscillator and for
the square well potential can be found in\cite{gr}.

Let us consider the case of the linear-rising potential, $v(r)=\gamma r$.
Notice that $\gamma|{\mbox{\boldmath $r$}}-\sigma\hat{\mbox{\boldmath $q$}}|-
\gamma|{\mbox{\boldmath $r$}}|\simeq -\gamma z\sigma /r$ for large $r$, and
therefore the integrand in Eq. (\ref{c6b}) does not increase with $r$.
(It reflects the elimination of the infra-red singularity in the expansion
(\ref{b13})).
Using this result we perform the $\sigma$-integration in
Eq. (\ref{c6b}) and then relate ${\cal F}_1$ to the zero order term
${\cal F}_0$ by
\begin{equation}
{\cal F}_1\simeq i\gamma\frac{{\cal A}_1}{4\pi}\frac{\partial^2}{\partial y^2}
\ \int_{-\infty}^{\infty}ds\exp (-iys)\int d^3r\frac{z}{r}
\phi({\mbox{\boldmath $r$}}-
s\hat{\mbox{\boldmath $q$}})\phi({\mbox{\boldmath $r$}})
\simeq \frac{\gamma{\cal A}_1}{2r_0}
\frac{\partial^3}{\partial y^3}{\cal F}_0(y)
\label{c7}
\end{equation}
where $r_0$ is some average size of the system which determines
the slope of the parton momentum distribution.
Approximating the
structure function ${\cal F}_0(y)$ at small $y$ by a Gaussian,
${\cal F}_0(y)\sim \exp (-r_0^2y^2)$, and using Eq. (\ref{c7})
one obtains the estimate of the first correction term
\begin{equation}
\frac{{\cal F}_1(y)}{Q^2{\cal F}_0(y)}\sim {\cal A}_1\frac{y\Delta}{Q^2}
\label{c8}
\end{equation}
where $\Delta =\gamma r_0$ is of the order of the first excitation energy.
It means that the correction term is indeed small, so that the approximation
of the structure function by ${\cal F}_0$ should be valid also for
{\em non-asymptotic} values of $Q^2\gtrsim$ few (GeV/c)$^2$.

\section{Comparison with data}

In Fig. \ref{fig3} we displayed the proton structure function $F_2^p(x,Q^2)$
from BCDMC\cite{bcd} and SLAC-MIT\cite{slac} experiments,
as functions of $Q^2$ for fixed $x$.
The solid curves correspond to the 15-parameter
fit to these data taken from\cite{nmc}:
\begin{equation}
F_2^p(x,Q^2)=A(x)\left [\frac{\ln (Q^2/\Lambda^2)}{\ln (Q^2_0/\Lambda^2)}
\right ]^{B(x)}\left (1+\frac{C(x)}{Q^2}\right ),
\label{d1}
\end{equation}
where $Q_0^2=20$ GeV$^2$, $\Lambda =250$ MeV, and
\begin{eqnarray}
A(x) = \frac{(1-x)^{2.562}}{x^{0.1011}}\sum_{\ell =0}^{4}
a_{\ell}(1-x)^{\ell}; ~
B(x) = 0.364-2.764x+\frac{0.015}{x+0.0186}; ~
C(x) = \sum_{\ell =1}^{4}c_{\ell}x^{\ell}\nonumber ,
\end{eqnarray}
where $\{ a_{\ell}\}$=(0.4121, -0.518, 5.967, -10.197, 4.685), and
$\{ c_{\ell}\}$=(-1.179, 8.24, -36.36, 47.76).

Let us treat this fit as the actual data and display
it in Fig. \ref{fig10}
as a function of the scaling variable $\bar x$, Eq. (\ref{b21}):
$\tilde F_2^p(\bar x,Q^2)=F_2^p[x(\bar x,Q^2),Q^2]$ (the dashed lines).
The spectator is taken to be a diquark of mass $m_s=m_d=850$ MeV.
This value of $m_d$ is taken from
the bag model or the non-relativistic quark model, which estimate
the scalar and vector diquark masses somewhere within the range of 600 to
1100 MeV\cite{thom1,thom2}. For a comparison we also displayed
in Fig. \ref{fig10}, the structure function $F_2^p(x,Q^2)$, Eq. (\ref{d1}),
as a function of $Q^2$ for fixed values of $x$ (the solid lines).
The values of $Q^2$ for $\bar x=$ const are taken
within the limits of the data, i.e. $x(\bar x,Q^2)\leq 0.75$. Therefore
the minimal values of $Q^2$ for $\bar x$=0.65 and $\bar x$=0.7 are smaller than
the corresponding values of $Q^2$ for $x$=0.65 and  $x$=0.7.
Since the values of $x$ and $Q^2$ for
$\bar x=0.75$ are outside the kinematical region of the BCDMC and SLAC-MIT
experiments ($x(\bar x,Q^2)>0.75$), the case of $x,\bar x =0.75$
is not shown in Fig. \ref{fig10}.

One observes that $\tilde F_2^p$ at constant $\bar x$
exhibits considerably weaker $Q^2$-dependence than the same structure
function taken at constant $x$. It implies that the essential part
of the Bjorken scaling violations,
usually attributed to the logarithmic terms,
is in fact $1/Q^2$-corrections, incorporated in the scaling
variable $\bar x$. In order to assess what part of the $x$-scaling violation is
accounted for by use of the scaling variable
$\bar x$, we consider the following procedure.
Let us take the proton structure function, $F_2^p(\bar x,Q^2)$,
at $Q^2=250$ (GeV/c)$^2$, which corresponds
to the largest value of $Q^2$ for the
data sets in Fig. \ref{fig3}. In that region of $Q^2$
the scaling variable $\bar x$ is very close to $x$, and therefore
$F_2^p(x,250)\cong \tilde F_2^p(x,250)$ (see Fig. \ref{fig10}).
The corresponding ``asymptotic"
proton structure function $f^p(x)=F_2^p(x,250)$, obtained from the fit
(\ref {d1}), is shown in Fig. \ref{fig11}, together with three data
points for $Q^2=250$ (GeV/c)$^2$. The dotted part corresponds to the
same fit, Eq. (\ref{d1}), extended beyond the limit of the data ($x > 0.75$).
The scaling of the structure function in the $\bar x$ variable,
$\tilde F_2^p(\bar x,Q^2)=f^p(\bar x)$, generates the $Q^2$-dependence
of the same structure structure function taken at constant $x$,
\begin{equation}
F_2^p(x,Q^2)=f^p[\bar x(x,Q^2)].
\label{d2}
\end{equation}
Then the deviations of Eq. (\ref{d2}) from the data
would explicitly show of what part of the $x$-scaling violations
is {\em not} incorporated in the variable $\bar x$.

The results of this analysis are presented in Fig. \ref{fig12},
with $F_2^p(x,Q^2)$, Eq. (\ref{d2}), given by dashed lines.
As in the previous analysis (Fig. \ref{fig10}) the spectator has
been taken to be a diquark, $m_s=m_d=850$ MeV and $C=0$ in
Eq. (\ref{b27}).
One sees that the $Q^2$-dependence of at $x=0.75$ is well
reproduced. The $Q^2$-dependence of the other data sets is reproduced
only partially, and the deviations from data increase for smaller $x$.
However, the increase of spectator mass, $m_s$,
as $1-x$, Eq. (\ref{b27}),
can well influence the $Q^2$-dependence even in the region of small
$x$. The evolution of the spectator mass with $x$ and $Q^2$ is given
by the coefficient $C$ in Eq. (\ref{b27}).
For large $x$ this coefficient can be taken as a constant.
It then may be extracted from data by requiring that
Eq. (\ref{d2}) reproduce two large $x$ data sets, for
instance $x=0.75$ and $x=0.65$. Since each data set is fitted by adjusting
the spectator mass $m_s$, Eq. (\ref{b27}) fixes also the parameter
$m_d$, which is the value of diquark mass
in the elastic limit $(x=1)$. Since in this limit the
nucleon is not excited, one expects
$m_d$ to be on the order of two constituent quark masses, i.e.
500 - 1000 MeV\cite{thom1,thom2,thom3}. It appears however, that
the two data sets cannot be fitted with such values of
$m_d$, but only with $m_d\approx 0$ and
$C\approx 3$ (GeV)$^2$. This corresponds to a spectator built out of
very light quarks.

Taking $m_d=0$ in Eq. (\ref{b27}), we find for the scaling variable
$\bar x$, Eq. (\ref{b21}),
\begin{equation}
\bar x=\frac{x+\sqrt{1+4M^2x^2/Q^2}-
\sqrt{(1-x)^2+4C(1-x)x^2/Q^2}}
{1+\sqrt{1+ 4M^2x^2/Q^2}}
\label{d3}
\end{equation}
One gets from Eq. (\ref{d3}) that
$\bar x=1$ for $x=1$, so that the two scaling variables vary
within the same limits.

Proton structure function given by Eqs. (\ref{d2}), (\ref{d3}) for $C=3$
is shown in Fig. \ref{fig13} by dashed lines.
Rather good agreement with the data is observed even for $x<0.65$,
although one expects large logarithmic corrections in this region and
also the variation of $C$ should be taken into account when
$x$ is far from 1. It is therefore of greater interest to make
a comparison for $x > 0.75$. We show in Fig. \ref{fig14}a the
data for the proton structure function taken
from new SLAC measurements in
the threshold region for $7<Q^2<30$ (GeV/c)$^2$\cite{thr},
together with three high-statistics spectra
for $Q^2$=5.9, 7.9, and 9.8 (GeV/c)$^2$ from
a previous SLAC experiment\cite{thr1}.
These data do not scale either in the variable $x$
(Fig. \ref{fig14}a), or in the Nachtmann variable $\xi$\cite{thr}.
In contrast, excellent scaling (Fig. \ref{fig14}b) is observed when
the data are plotted as a function of
$\bar x$, Eq. (\ref{d3}), with $C$=3. It is even more remarkable that the
structure function obtained from these data completely coincides
with the asymptotic structure function $f^p(x)=F_2^p(x,250)$
taken from BCDMC experiment (Fig. \ref{fig11}) and
shown by the solid line.
This confirms the dominance of the quasi-free term ${\cal F}_0$,
Eq. (\ref{b13}), which is in line with our estimates in the previous Section.
Indeed, the lowest correction term, ${\cal F}_1$,
is proportional to $y$, Eq. (\ref{c8}).
Since $y=0$ for $x=1$ and $m_s=0$, Eq. (\ref{b19}), it follows that
${\cal F}_1\to 0$ near the elastic threshold.

The analysis of the large $x$ data\cite{thr,thr1} (Fig. \ref{fig14}b)
allows us to extract the
asymptotic structure function $f^p(\bar x)=\tilde F_2^p(\bar x, Q^2)$,
up to $x \lesssim 0.95$. We find that $f^p(\bar x)$ is clearly
below the dashed curve for $x\gtrsim 0.8$, which is the fit (\ref{d1}),
extended outside the data, Fig. \ref{fig11}. It means that the fit
(\ref{d1}) is not applicable in that region.
It was shown by Drell, Yan, and West\cite{drell,west}
that the threshold ($x\to 1$) behaviour
of the asymptotic structure function is correlated with the large
$Q^2$ behaviour of the elastic form factor. Using
quark counting arguments one obtains $f^p(x)\sim (1-x)^3$.
However additional QCD effects may effectively
increase the value of the exponent\cite{brod}.
This predictions can be checked by a comparison with the structure
function that we extracted from the data. We find that
it well described by $f^p(x)=1.5(1-x)^{3.7}$
for $x\gtrsim 0.75$ (the dot-dashed curve in Fig. \ref{fig14}b).

With the asymptotic structure function $f^p(\bar x)$,
extracted from the experiment\cite{bcd,thr,thr1},
we can predict the structure function at large values of $x$
from moderate up to very high values of $Q^2$. The results are
shown in Fig. \ref{fig15}. The dashed lines correspond to Eq. (\ref{d2}),
with the asymptotic structure function $f^p(x)$ given by the fit
(\ref{d1}) for $x\leq .78$, $Q^2=$ 250 (GeV/c)$^2$, and
$f^p(x)=1.5(1-x)^{3.7}$ for $x\geq 0.78$. We plot also a few available
data points, taken from old SLAC-MIT measurements\cite{atw},
for $x=$ 0.78, 0.82 and 0.86.
Again our predictions are in full agreement with the data. Still
a check of our predictions for higher $Q^2$ would be of greater
interest, since we predict a significant non-logarithmic $Q^2$-dependence
for large $x$, Fig. \ref{fig15}. In fact, as follows from our analysis,
none of the {\em large} $x$ data ($x\gtrsim 0.5$) exhibit any substantial
deviations from the $\bar x$-scaling. Therefore, measurements for
high $Q^2$ would be extremely important in order to establish
the importance of logarithmic corrections in the large $x$ region.

\section{Summary}

In this paper we concentrated on effects of confinement
in deep inelastic scattering.
Using the framework of the Bethe-Salpeter equation, we found a new
expansion of the structure function in powers of $1/Q^2$ that
is free of infra-red singularities and diminishes corrections
to the zero-order term. The zero-order term describes scattering off a free
off-mass-shell parton, which keeps the same off-shell mass
in the final state. We evaluated corrections from higher
order terms for the case of a linear-rising confining potential,
and found them small even for rather low values of $Q^2$.
It allows us to consider the zero order (quasi-free) term as a good
approximation for the structure function, valid in the entire $Q^2$ region.

By analyzing the quasi-free term we found it depends on the
scaling variable $\bar x$. This variable coincides with the Bjorken
variable $x$ in the limit $Q^2\to\infty$. However, at finite
$Q^2$ the variables $x$ and $\bar x$ are quite different:
$x-\bar x \sim m_s^2/(1-x)Q^2$, where $m_s$ is the invariant mass
of the spectator particles. It implies that $1/Q^2$ corrections to
the $x$ scaling would be much larger than those obtained in perturbative
calculations, especially in the region of large $x$. These corrections
could be very appreciable also at small $x$, since the spectator
mass increases with $(1-x)$ due to gluons emission. However the evaluation
of the spectator mass at small $x$ depends on knowledge of
its $Q^2$-dependence, which
can be obtained by using the evolution equation. In this paper
we limited our analysis to the large $x$ region, where the
$Q^2$-dependence of the spectator mass $m_s$ is less important.

Using simple arguments we showed that $m_s^2=m_d^2+C(1-x)$
for $x\lesssim 1$. Here $m_d$ is the diquark mass, since there is no
gluon emission at the elastic threshold. First we analysed
the data for the proton structure function by taking for
$m_d$ values between 500-1000 MeV, i.e., on the order of two
constituent quark masses, and taking $C=0$.
Even though we found that scaling
in $\bar x$ is certainly better than that in $x$,
the scaling deviations are still considerable, especially
when $\bar x$ approaches the elastic threshold.
Next we included the gluon emission contribution to the spectator mass
by taking $C\not =0$. However, instead of taking for the diquark mass, $m_d$,
the values from constituent quark models, we considered the nucleon
structure function data as a source of information for the value of
$m_d$. We found that
{\em all} large $x$ data ($x\gtrsim 0.5$) display perfect scaling
in the $\bar x$-variable for $C\approx 3$ and $m_d\approx 0$,
which corresponds to very light quarks.

Since $\bar x\to x$ for $Q^2\to\infty$, the perfect scaling in the
$\bar x$-variable allows us to arrive at the Bjorken limit already
at moderate values of $Q^2$. Thus our analysis of the proton structure
function near elastic threshold shows that $F_2\sim (1-x)^{3.7}$ for
$x\to 1$. This is different from the theoretical results based on
simple quark counting arguments, which predict $F_2\sim (1-x)^{3}$.

Untill now the measurements of the structure function for large values of $x$
($x>0.75$) have not been extended beyond $Q^2\sim 25$ (GeV/c)$^2$. The present
large $x$ data are in full agreement with our predictions
based on $\bar x$-scaling, and thus do not display
any noticeable effects of logarithmic terms. It would be very
interesting to extend the measurements to higher values of $Q^2$.
The $\bar x$-scaling predicts a distinctive $Q^2$-dependence
in the structure function $F(x,Q^2)$ at large fixed
values of the Bjorken variable $x$. Hence,
the deviations from our predictions would establish the
magnitude of the logarithmic scaling violations at large $x$.

\section{Acknowledgments}

I am  grateful to A. Goldhaber and V. Zakharov
for useful discussions. Special thanks to A. Rinat and B. Svetitsky for
reading the manuscript and making valuable comments on it.

\begin{figure}[tb]
\hspace*{\fill}
\hspace*{\fill}
\caption[]{Nucleon structure function given by the imaginary part of the
forward Compton amplitude. The first diagram is the Impulse Approximation, and
the second one describes the Final State Interaction. The shaded area includes
spectator particles (quarks and gluons).
\label{fig1}}
\end{figure}

\begin{figure}[tb]
\caption[]{Nucleon vertex function, which describes quark and gluon emission.
Quarks are shown by solid lines and gluons by wavy lines.
\label{fig2}}
\end{figure}

\begin{figure}[tb]
\caption[]{Proton structure function from BCDMS\cite{bcd} and
SLAC-MIT\cite{slac}  experiments. The solid curves correspond to a
15 parameter fit to these data, taken from\cite{nmc}.
\label{fig3}}
\end{figure}

\begin{figure}[tb]
\caption[]{Bethe-Salpeter equation for the operator $T$ describing
interactions between the struck parton and the spectator partons in the
final state.
\label{fig4}}
\end{figure}

\begin{figure}[tb]
\caption[]{Bethe-Salpeter equation for the vertex $\Gamma$
describing the relativistic bound state.
\label{fig5}}
\end{figure}

\begin{figure}[tb]
\caption[]{Diagrammatic representation of the first two terms
of the expansion (\ref{b13}). Modified propagator of the struck parton
is marked by ``$\sim$".
\label{fig6}}
\end{figure}

\begin{figure}[tb]
\caption[]{Diagrammatic representation of the leading term.
Quarks and gluons are shown by solid and wavy lines respectively.
The modified propagators are marked by ``$\sim$".
\label{fig7}}
\end{figure}

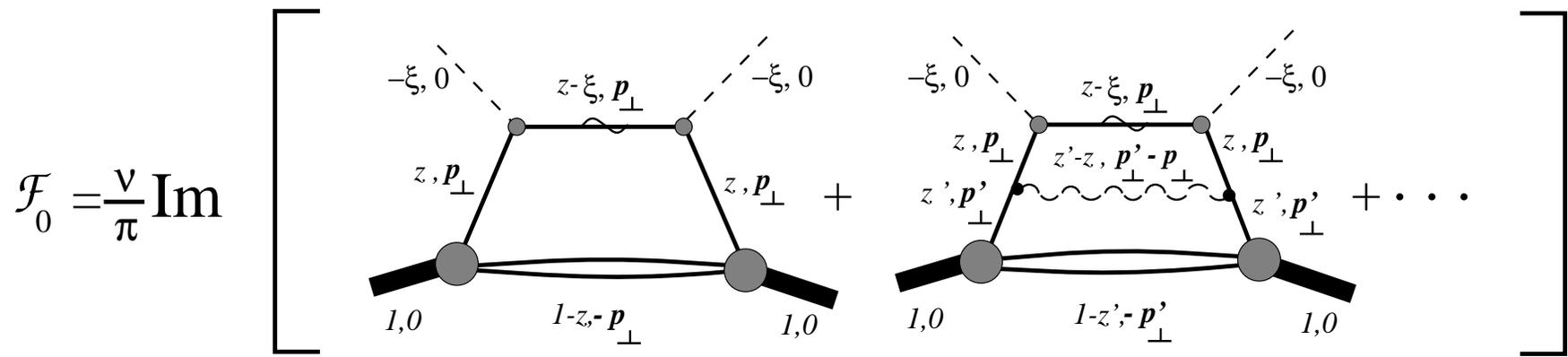
\begin{figure}[tb]
\caption[]{The same as in Fig. \ref{fig7}, but using light-cone variables.
The negative $z$-axis has been chosen along ${\mbox{\boldmath $q$}}$.
\label{fig8}}
\end{figure}

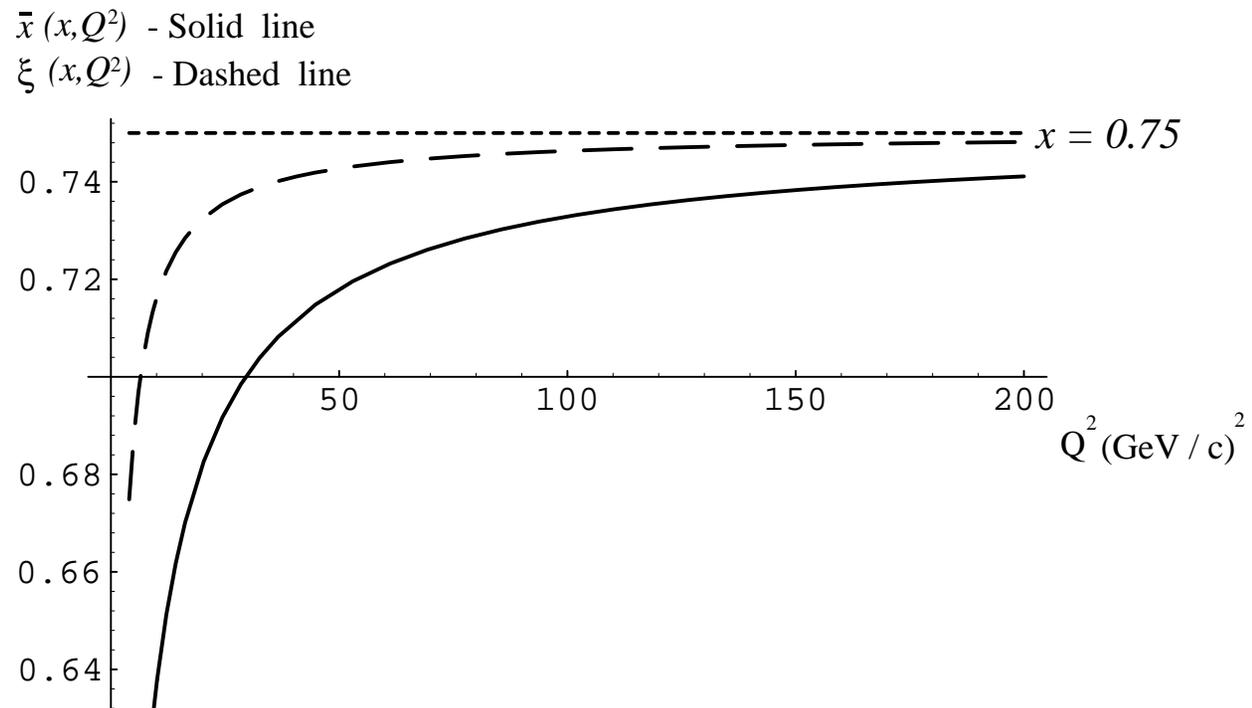
\begin{figure}[tb]
\caption[]{Comparison between the scaling variable $\bar x\equiv \bar x(x,Q^2)$
and the Nachtmann variable $\xi\equiv \xi (x,Q^2)$,
Eqs. (\ref{b21}), (\ref{b22}) for $x=0.75$. The spectator mass
equals the nucleon mass.
The dotted line corresponds to $x$=0.75.
\label{fig9}}
\end{figure}

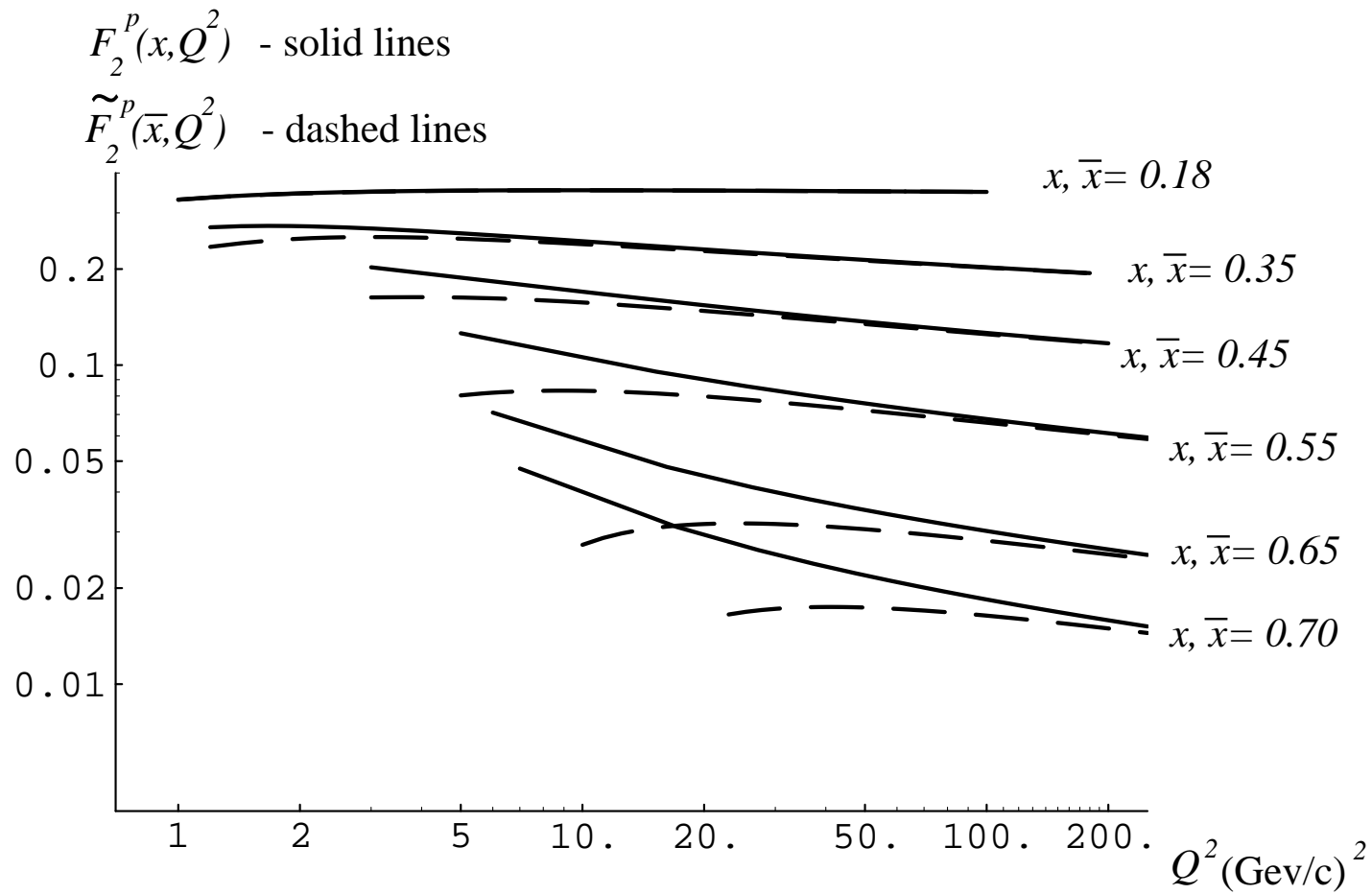
\begin{figure}[tb]
\caption[]{The proton structure function plotted as a function of $Q^2$
for fixed values of the Bjorken variable $x$ (solid lines) and the
scaling variable $\bar x$, Eq. (\ref{b21}) (dashed lines).
\label{fig10}}
\end{figure}

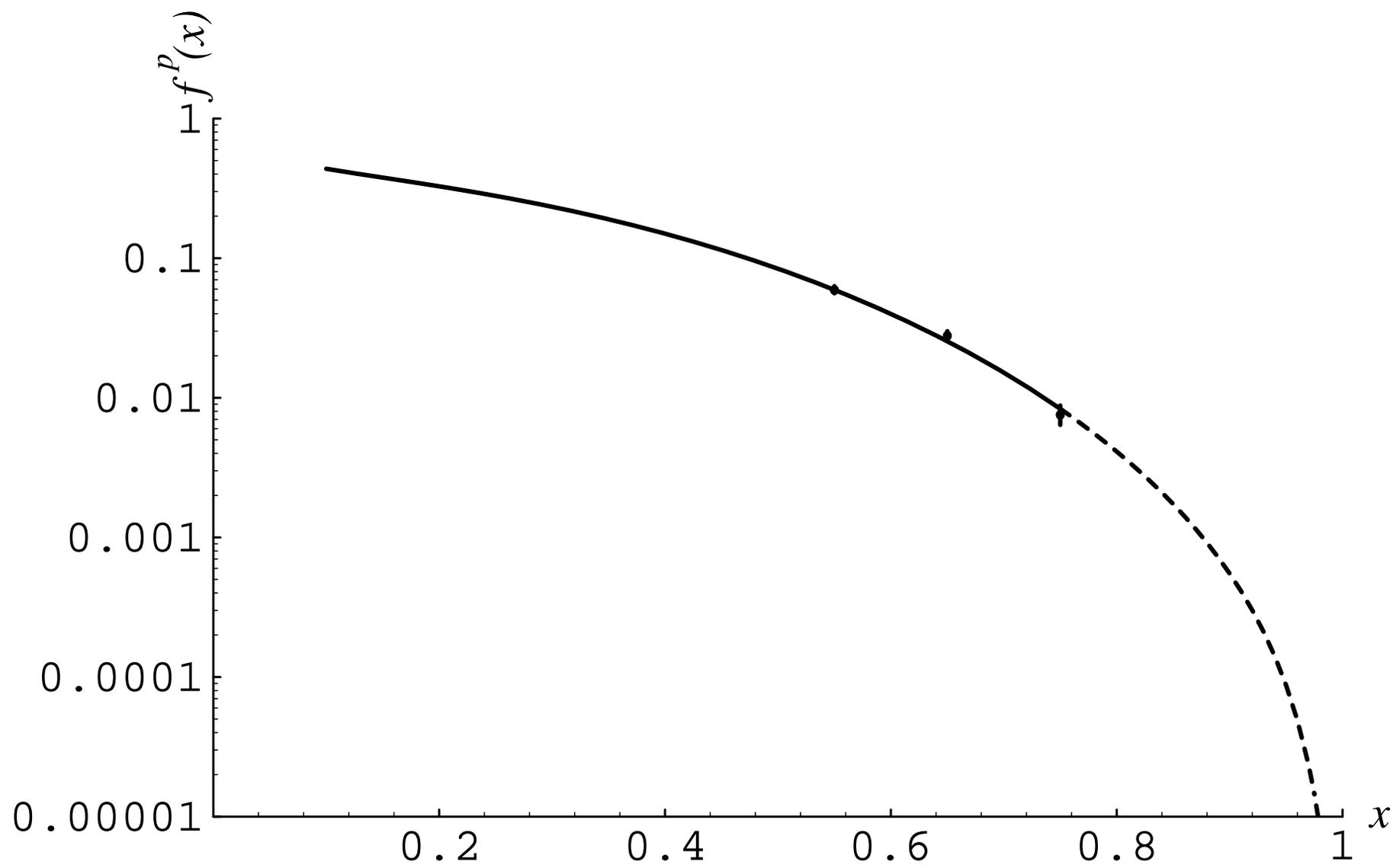
\begin{figure}[tb]
\caption[]{Proton structure function in the asymptotic region,
$f^p(x)=F^p_2(x,250)$,
given by the fit, Eq. (\ref{d1}). Three data points are taken from\cite{bcd}.
The dashed part of the curve lies in
the region outside the data\cite{bcd}.
\label{fig11}}
\end{figure}

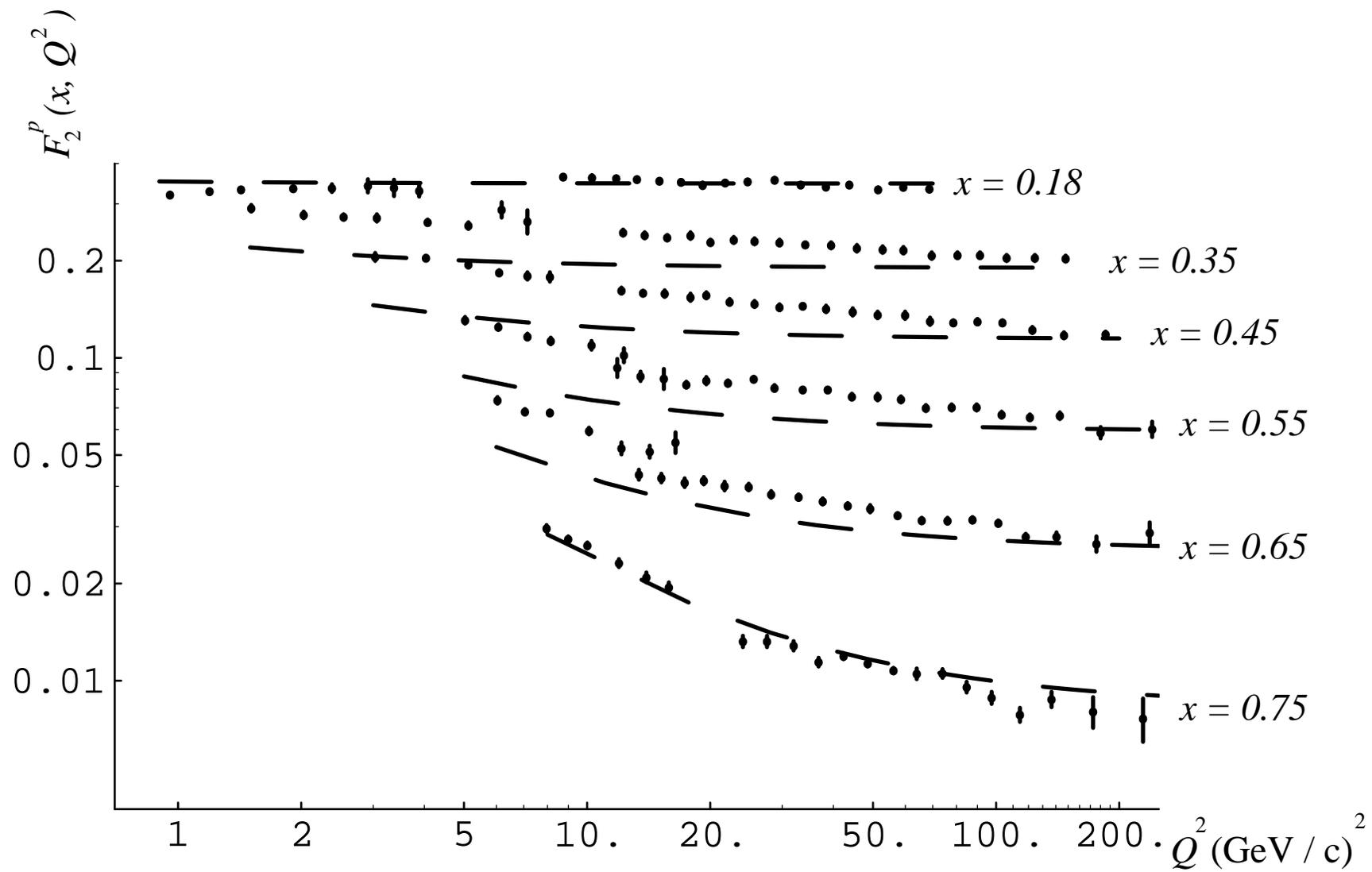
\begin{figure}[tb]
\caption[]{$Q^2$-dependence of the structure function $F_2^p(x,Q^2)$,
which corresponds to scaling in
the variable $\bar x$, Eq. (\ref{d2}).
The spectator is taken to be a diquark of mass $m_s=m_d$=850 MeV.
\label{fig12}}
\end{figure}

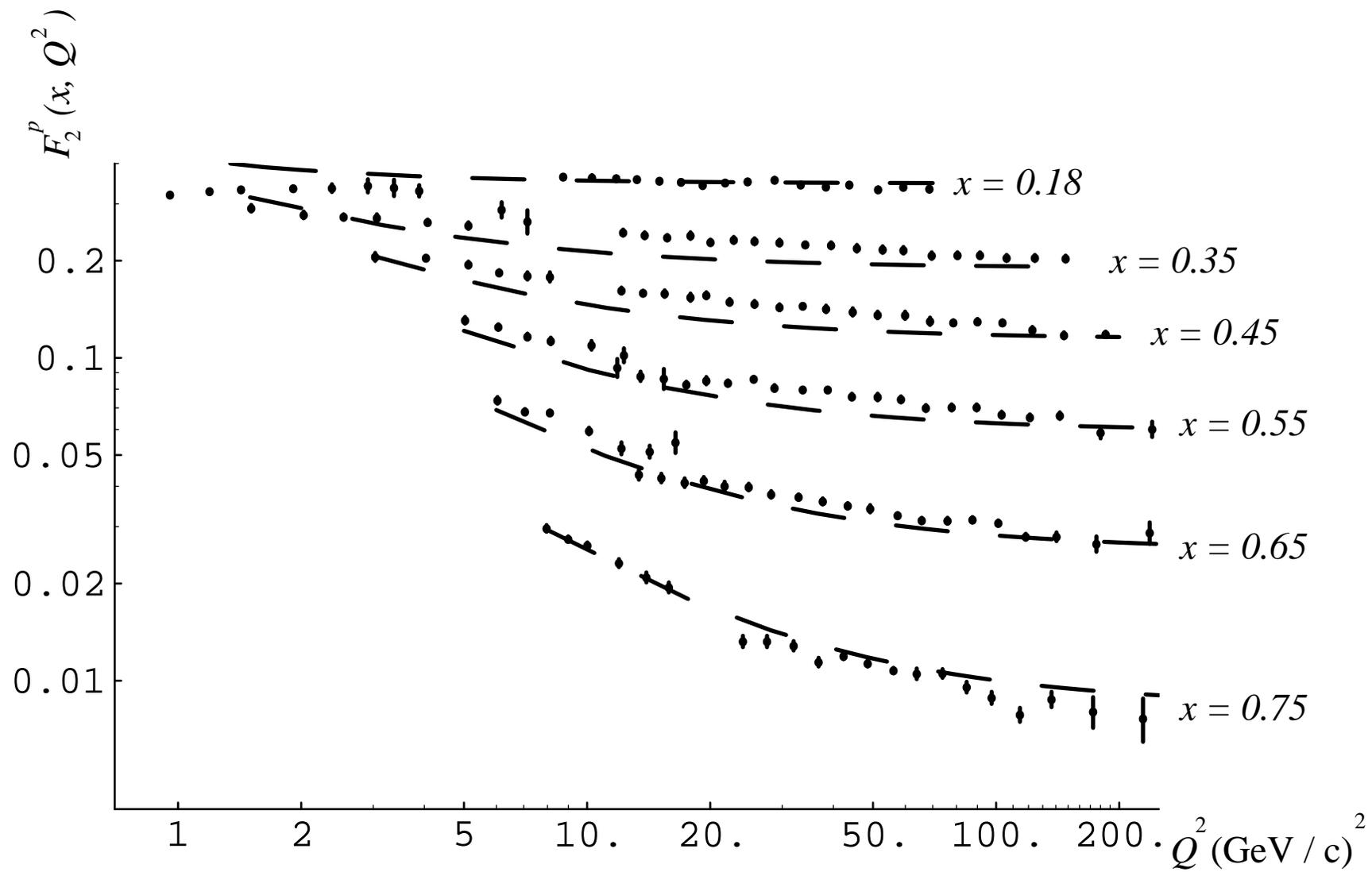
\begin{figure}[tb]
\caption[]{The same as in Fig. \ref{fig12}, but where the spectator mass is a
function of $x$, Eq. (\ref{b27}), for $m_d=0$ and $C=3$ (GeV)$^2$.
\label{fig13}}
\end{figure}

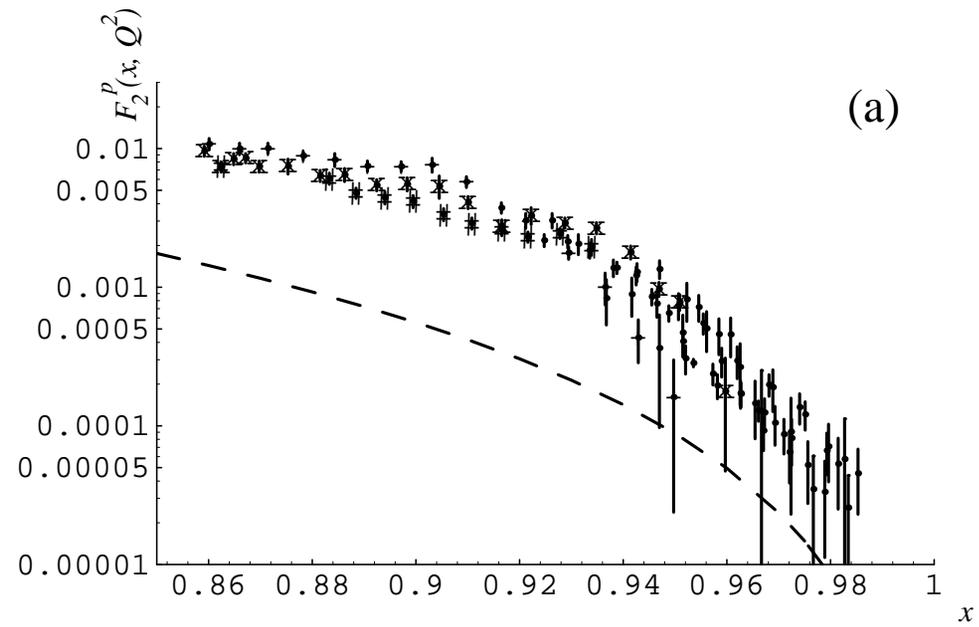
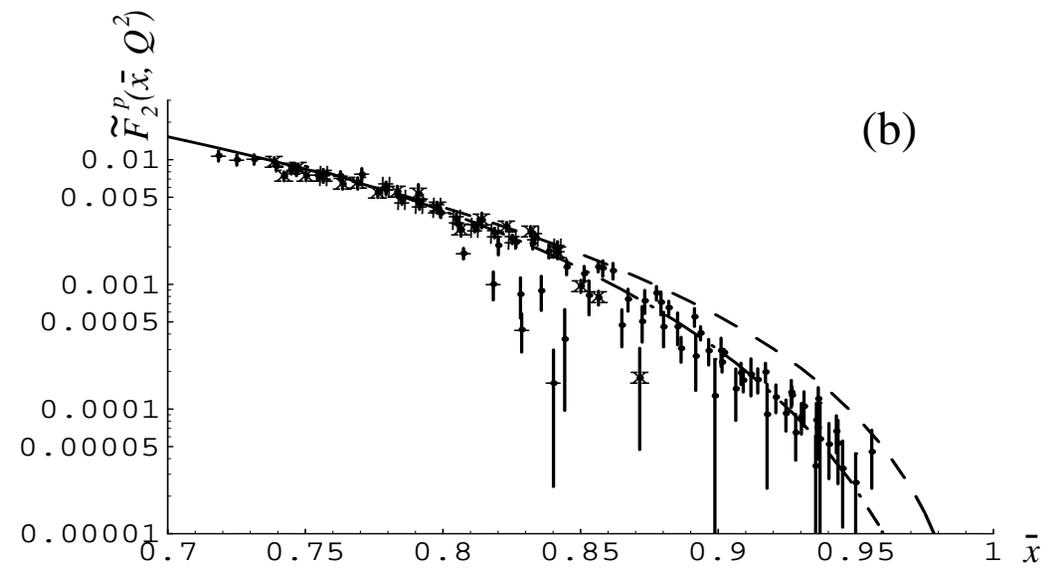
\begin{figure}[tb]
\caption[]{The structure function in the region of large $x$
for $7<Q^2<30$ (GeV/c)$^2$\cite{thr,thr1}, plotted (a) as a function of
the scaling variable $x$ and (b) as a function of the scaling variable
$\bar x$, Eq. (\ref{d3}). Three high-statistics data sets\cite{thr1}
for $Q^2$=5.9, 7.9, and 9.8 (GeV/c)$^2$ are marked by
``+", ``x", and ``{\#}" respectively. The solid curve and the dashed
curves correspond to the asymptotic structure function, shown in
Fig. \ref{fig11}. The dot-dashed curve in (b) corresponds to
$\tilde F_2^p(\bar x,Q^2)=1.5(1-\bar x)^{3.7}$.
\label{fig14}}
\end{figure}

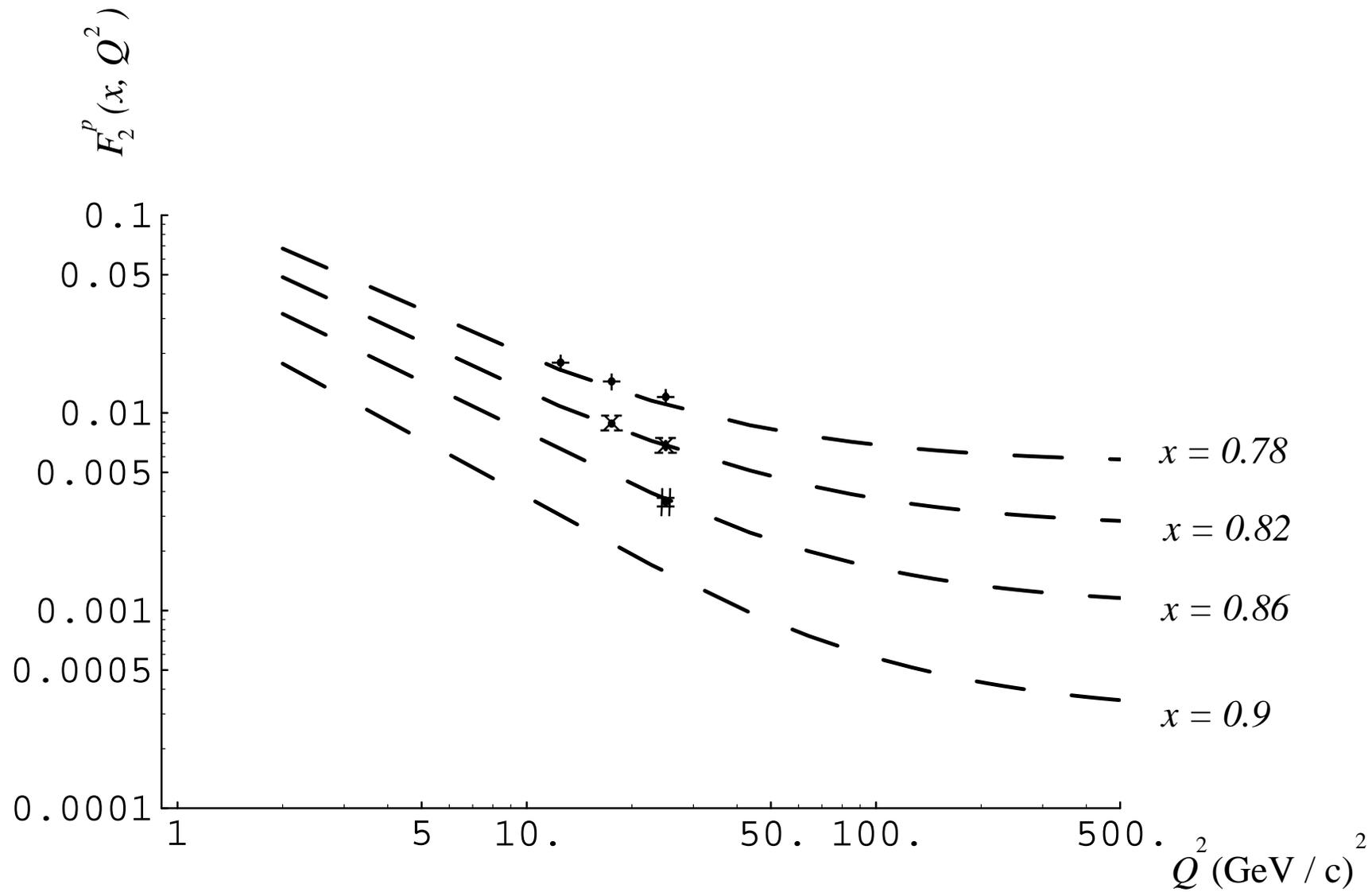
\begin{figure}[tb]
\caption[]{Predictions for the structure function $F_2^p(x,Q^2)$
in the region of large $x$. The data points\cite{atw}
correspond to $x=$0.78 (+), 0.82 (x), 0.86 ({\#}).
\label{fig15}}
\end{figure}

\end{document}